\documentstyle[twoside,fleqn,espcrc2]{article}

\newcommand{\nn}{\nonumber}

\newcommand{\ovl}[1]{\overline{#1}}
\newcommand{\wt}[1]{\widetilde{#1}}

\newcommand{\p}{\partial}

\newcommand{\lslash}{l\kern-1ex /}
\newcommand{\pslash}{p\kern-1ex /}
\newcommand{\Dslash}{{\cal D}\kern-1.5ex /}
\newcommand{\bpsi}{\overline{\psi}}

\title{
\hfill\begin{minipage}{0pt}\scriptsize\vspace*{-1.5cm} \begin{tabbing}
\hspace*{\fill} UTHEP-388\\ 
\hspace*{\fill} UTCCP-P-48 
\end{tabbing} 
\end{minipage}\\[-8pt]
One-loop renormalization factors and mixing coeffecients 
of bilinear quark operators
for improved gluon and quark actions
\thanks{Talk presented by Y.~Taniguchi}}

\author{Sinya Aoki $^{\rm a}$, Kei-ichi Nagai $^{\rm b}$,
Yusuke Taniguchi
\address{Institute of Physics, University of Tsukuba, 
Tsukuba, Ibaraki-305, Japan}
and Akira Ukawa $^{\rm a}$
\address{Center for Computational Physics, University of Tsukuba,
Tsukuba, Ibaraki-305, Japan}}

\begin{document}

\begin{abstract}
We calculate one-loop renormalization factors and mixing coefficients
of bilinear quark operators 
for a class of gluon actions with six-link loops and O(a)-improved
quark action.
The calculation is carried out by evaluating on-shell Green's functions
of quarks and gluons in the standard perturbation theory.
We find a general trend that finite parts of one-loop
coefficients are reduced approximately by a factor two for the
renormalization-group improved gluon actions
compared with the case of the standard plaquette gluon action.
\end{abstract}

\maketitle

\section{Introduction}

In numerical studies of field theories through computer simulations
finite lattice spacing effects poses an obstacle in extracting physical 
predictions for observable quantities. 
In an effort to reduce this problem, a recent full QCD 
simulation by the CP-PACS Collaboration\cite{CPPACS-production} employs 
a combination of the $O(a)$-improved clover quark action\cite{SW85} and
a renormalization-group-improved gluon action with 6-link loops
\cite{Iwasaki83},  based on previous results of a 
comparative study of various action combinations\cite{CPPACS-comparison}. 
With this development it has become necessary to calculate renormalization 
factors and mixing coefficients of quark operators for improved gluon 
and quark actions.  We have carried out such a calculation to one-loop 
order of perturbation theory for 
bilinear quark operators~\cite{improved-Z,improved-bc}.
In this article we summarize the results. 

Our calculations are performed within standard perturbation theory for
on-shell Green's functions of quarks and gluons. 
For the plaquette gluon action, perturbative calculation of the mixing 
coefficients of bilinear quark operators has been previously carried out  
using the Schro\"dinger functional technique\cite{Luescher96,Sint-Weisz}. 
A recent work with the same technique\cite{Aoki} extends the results to 
the same set of improved gluon actions as examined by us.  
We find a good agreement between results by us and those reported in 
these references.

\begin{table*}[hbt]
\setlength{\tabcolsep}{0.15pc}
\newlength{\digitwidth} \settowidth{\digitwidth}{\rm 0}
\catcode`?=\active \def?{\kern\digitwidth}
\caption{Finite part $z_\Gamma$ of renormalization factor
for bilinear operator of form $(1+ma)\ovl{\psi}\Gamma \psi$ 
({\it i.e.,} $z=0$ in (\protect\ref{eq:bilinear})) with $m$ the 
subtracted quark mass incorporating one-loop self-energy correction.
Coefficients of the term $c_{SW}^n (n=0,1,2)$ in an expansion 
$z_\Gamma=z_\Gamma^{(0)}+c_{SW}z_\Gamma^{(1)}+c_{SW}^2z_\Gamma^{(2)}$ 
are given in the column labeled $(n)$. Errors are at most in the last 
digit given.}
\label{tbl:zfact}
\begin{tabular*}{\textwidth}{ll|lll|lll|lll|lll}
\hline
\multicolumn{2}{c|}{gauge action}&
\multicolumn{3}{c|}{$V$} &
\multicolumn{3}{c|}{$A$} &
\multicolumn{3}{c|}{$S$} &
\multicolumn{3}{c}{$P$} \\
$c_1$ & $c_{23}$ & 
(0) & (1) & (2)  &
(0) & (1) & (2)  &
(0) & (1) & (2)  &
(0) & (1) & (2)  \\
\hline
$  0    $&$0    $&$-20.618$&$4.745$&$0.543$&$-15.797$&$-0.248$&$2.251$&
$-12.953$&$-7.738$&$1.380$&$-22.596$&$2.249$&$-2.036$\\
$ -1/12 $&$0    $&$-16.603$&$4.228$&$0.464$&$-12.540$&$-0.198$&$2.021$&
$ -9.607$&$-6.836$&$1.367$&$-17.734$&$2.015$&$-1.745$\\
$ -0.331$&$0    $&$-11.099$&$3.326$&$0.336$&$ -8.192$&$-0.125$&$1.610$&
$ -4.858$&$-5.301$&$1.266$&$-10.673$&$1.601$&$-1.281$\\
$ -0.27 $&$-0.04$&$-11.540$&$3.418$&$0.353$&$ -8.523$&$-0.131$&$1.657$&
$ -5.260$&$-5.454$&$1.292$&$-11.292$&$1.644$&$-1.316$\\
$ -0.252$&$-0.17$&$-10.525$&$3.248$&$0.338$&$ -7.707$&$-0.117$&$1.587$&
$ -4.366$&$-5.166$&$1.287$&$-10.001$&$1.565$&$-1.212$\\
\hline
\end{tabular*}
\end{table*}

\section{Choice of actions}

We adopt the clover action for quark,
which is given by adding the clover term
\begin{eqnarray}
S_{\rm C} = - c_{\rm SW} a^3 \sum_n \sum_{\mu, \nu}
ig \frac{r}{4} \bpsi_n \sigma_{\mu \nu} P_{\mu \nu} (n) \psi_n
\nn
\end{eqnarray}
to the ordinary Wilson fermion action.
The gluon action we consider is defined by 
\begin{eqnarray}
S_{\rm gluon} = \frac{1}{g^2} {\rm Tr}\left\{
c_0 U_{pl}+c_1 U_{rtg}+c_2 U_{chr}+c_3 U_{plg}\right\}, 
\nn
\end{eqnarray}
where the first term represents the standard plaquette term, and the 
remaining terms are six-link loops formed by a $1\times 2$ rectangle, 
a bent $1\times 2$ rectangle (chair) and a 3-dimensional parallelogram. 
The coefficients $c_0, \cdots, c_3$ satisfy
$c_0+8c_1+16c_2+8c_3=1$. 
At the one-loop level, the choice of the gluon action is specified by the
pair of numbers $c_1$ and $c_{23}=c_2+c_3$.
We adopt the following five cases in our calculation:
(i) the standard plaquette action $c_1=0, c_{23}=0$, 
(ii) the tree-level improved action in the Symanzik approach
$c_1=-1/12, c_{23}=0$\cite{Weisz83},
and (iii) three choices suggested
by an approximate renormalization-group analysis, $c_1=-0.331, c_{23}=0$
and $c_1=-0.27, c_{23}=-0.04$ by Iwasaki\cite{Iwasaki83},
and $c_1=-0.252, c_{23}=-0.17$ by Wilson\cite{Wilson80}.

\section{Bilinear operators and  improvement}

We consider bilinear quark operators of form, 
\begin{equation}
{\cal O}^\Gamma = \bpsi \Gamma \psi,\qquad
\Gamma= 1, \gamma_5, \gamma_\mu, \gamma_\mu \gamma_5, \sigma_{\mu \nu}.
\label{eq:bilinear}
\end{equation}
At the tree-level, there exists a one-parameter family of 
${\cal O}(a)$-improved operators given by\cite{Heatlie91}
\begin{eqnarray}
{\cal O}_0^\Gamma
&=& \left[ 1+ar \left( 1-z \right) m_0 \right] \bpsi \Gamma \psi\nn\\
&&+ z \bpsi \Gamma^\otimes \psi
- z^2 \bpsi \Gamma ' \psi,
\label{eqn:improved-onshell}
\end{eqnarray}
where $\Gamma^\otimes$ and $\Gamma '$ are
${\cal O}(a)$ and ${\cal O}(a^2)$ vertices, 
and $m_0$ is the bare quark mass. 

At the one-loop level, on-shell Green's functions of these operators 
do not have terms of ${\cal O}(g^2 a \log a)$ 
when we set the clover term coefficient $c_{\rm SW} =1$ \cite{Heatlie91}.
The ${\cal O}(g^2 a)$ terms still remain, however.
In order to extract the renormalized operators without
${\cal O}(g^2 a)$ errors
we need to calculate mixing coefficients  $B_\Gamma$
and $C_\Gamma$ as well as the renormalization factor $Z_\Gamma$ defined by
\begin{eqnarray}
{\cal O}^\Gamma_{0} =
Z_\Gamma^{-1} {\cal O}^\Gamma_{\rm R}
- g^2 C_F a m_R\, B_\Gamma\, {\cal O}^\Gamma_{\rm R}
\nn\\
- g^2 C_F a\, C_\Gamma\, \wt{{\cal O}}^{\Gamma}_{\rm R} ,
\label{eqn:renormalization}
\end{eqnarray}
where $C_F$ denotes the second-order Casimir eigenvalue for the quark field, 
and the last two terms are needed to remove ${\cal O}(g^2 a)$ errors 
from on-shell matrix elements, with $\wt{{\cal O}}^{\Gamma}_{\rm R}$ 
a dimension 4 operator with derivative, {\it e.g., }
$\wt{{\cal O}}^A = \p_\mu {\cal O}^P$ and  
$\wt{{\cal O}}^V = \p_\mu {\cal O}^T$.

\begin{table*}[hbt]
\setlength{\tabcolsep}{0.6pc}
\caption{
Mixing coefficients 
for axial vector, vector, pseudo scalar and scalar density
at $z=0$.
}
\label{tbl:cb}
\begin{tabular*}{\textwidth}{ll|l|l|l|l|l|l}
\hline
\multicolumn{2}{c|}{gauge action}&
$C_A$ & $C_V$ & {$B_A$} & {$B_V$} & $B_P$ & $B_S$ \\
$c_1$ & $c_{23}$ &  &  &  &  &  &  \\
\hline
$0      $&$ 0     $&$ -0.005680(2)$&$-0.01226(3)$&
$0.1141(1)$&$ 0.1150(2)$&$ 0.1148(1)$&$ 0.1444(2)$\\
$-1/12  $&$ 0     $&$ -0.00451(1) $&$-0.01030(4)$&
$0.0881(1)$&$ 0.0886(2)$&$ 0.0890(1)$&$ 0.1144(2)$\\
$-0.331 $&$ 0     $&$ -0.00285(1) $&$-0.00729(4)$&
$0.0547(1)$&$ 0.0550(2)$&$ 0.0561(1)$&$ 0.0747(2)$\\
$-0.27  $&$ -0.04 $&$ -0.00302(1) $&$-0.00757(4)$&
$0.0572(1)$&$ 0.0575(2)$&$ 0.0586(1)$&$ 0.0777(2)$\\
$-0.252 $&$ -0.17 $&$ -0.00281(1) $&$-0.00705(4)$&
$0.0512(1)$&$ 0.0514(2)$&$ 0.0527(5)$&$ 0.0706(2)$\\
\hline
\end{tabular*}
\end{table*}

\section{Strategy of one-loop calculation}

We apply standard perturbation theory to the Green's functions
of the operator ${\cal O}_0^\Gamma$ and two on-shell external quarks 
to calculate renormalization factors and mixing 
coefficients.  In order to isolate terms of ${\cal O}(g^2a)$, one-loop 
amplitudes are expanded in powers of $a$, or equivalently in 
the external momenta $p$ and quark mass $m$.  
Infrared divergences that generally appear in on-shell amplitudes 
are regularized by a gluon mass $\lambda$. 

For this procedure to work, infrared divergences that contribute to 
the mixing coefficients $B_\Gamma$ and $C_\Gamma$ have to cancel out 
among diagrams. We have explicitly checked that this is in fact the 
case; infrared divergences appear only in the ordinary renormalization factor
$Z_\Gamma$, which, however, cancel against those in the renormalization 
factor for the continuum operator.  As a result, 
the renormalization factor in the $\ovl{\rm MS}$ scheme takes the form,
\begin{eqnarray}
Z_\Gamma^{-1} = 1+ \frac{g^2 C_F}{16\pi^2}
 \left( \left(\frac{h_2(\Gamma)}{4}-1\right)
\log (\mu a)^2 + z_\Gamma \right),
\nn
\end{eqnarray}
where $h_2(\Gamma)$ is an integer given by
$h_2(\Gamma)=4(A), 4(V), 16(P), 16(S), 0(T)$.

We note that the bare quark mass $m_0$ in (\ref{eqn:improved-onshell})
may be replaced by a subtracted mass $m=m_0-g^2C_F\Sigma_0/a$ incorporating 
the one-loop self-energy correction  
by a redefinition of  $z_\Gamma$, $B_\Gamma$ and $C_\Gamma$.  
We adopt this definition for numerical results
presented below. 

\subsection{Results}
 
The calculational procedure described above yields 
$z_\Gamma$, $B_\Gamma$ and $C_\Gamma$
as sums of one-loop integral constants.
While straightforward in principle, the actual algebra to evaluate the 
integrands is extremely tedious. 
We employ {\it Mathematica} to carry out this task, 
and also to generate the FORTRAN code from the results.
Infrared divergences that can appear in individual integrals are 
subtracted by working out the leading term of integrands 
for small loop momenta. 
Integrals are then evaluated by the Monte Carlo 
routine VEGAS in double precision.  
We generally employ $20$ sets of $10^5$ points for integration, 
except for $C_A$ for the plaquette action 
for which 20 sets of $10^6$ points are used.
Errors are estimated from variation of integrated values over the sets.  

We present our results for $z_\Gamma$, $C_\Gamma$ and $B_\Gamma$ 
for the case of $z=0$ in Tables~\ref{tbl:zfact} and \ref{tbl:cb}.
Results for $z\ne 0$ can be found in Refs.~\cite{improved-Z,improved-bc}.  
The Wilson parameter is taken to be $r=1$. 
For $z_\Gamma$, the coefficients of the expansion 
$z_\Gamma=z_\Gamma^{(0)}+c_{SW}z_\Gamma^{(1)}+c_{SW}^2z_\Gamma^{(2)}$ 
are given (Table~\ref{tbl:zfact}), while  
the constants $C_\Gamma$ and $B_\Gamma$ are evaluated for $c_{\rm SW} =1$
(Table~\ref{tbl:cb}).

We observe in the results that the renormalization factors and mixing 
coefficients are reduced by about a factor two for renormalization-group 
improved gluon actions compared to those for the plaquette action. 
More generally we find that $z_\Gamma$'s monotonically decrease
as $c_1$ and $c_{23}$ become large and negative\cite{improved-Z}.

\vspace*{3mm}
Numerical calculations have been carried out at Center for
Computational Physics, University of Tsukuba and at Research Institute for 
Fundamental Physics, Kyoto University. This work is supported in part by the
Grants-in-Aid of the Ministry of Education (Nos. 2373, 09304029,10640246).
KN and YT are JSPS Research Fellows.

\newcommand{\J}[4]{{\it #1} {\bf #2} (19#3) #4}
\newcommand{\MPL}{Mod.~Phys.~Lett.}
\newcommand{\IJMP}{Int.~J.~Mod.~Phys.}
\newcommand{\NP}{Nucl.~Phys.}
\newcommand{\PL}{Phys.~Lett.}
\newcommand{\PR}{Phys.~Rev.}
\newcommand{\PRL}{Phys.~Rev.~Lett.}
\newcommand{\AP}{Ann.~Phys.}
\newcommand{\CMP}{Commun.~Math.~Phys.}
\newcommand{\PTP}{Prog. Theor. Phys.}
\newcommand{\Suppl}{Prog. Theor. Phys. Suppl.}
\newcommand{\etal}{{\it et al.}}

\end{document}